\documentclass[fleqn,twoside,twocolumn,nofootinbib]{revtex4} 
\usepackage{ujp} 

\def\Journal#1#2#3#4{{#1} {\bf #2}, #4 (#3)}
\def\book#1#2#3#4{{#1}, {\it #2} (#3)#4}

\def\NPB{{Nucl. Phys.} B}
\def\PLB{{Phys. Lett.}  B}
\def\PRL{Phys. Rev. Lett.}
\def\PRD{{Phys. Rev.} D}

\def\PRP{Phys. Rep.}
\def\JETPl{JETP Lett.}
\def\JETP{Zh. Eksp. Teor. Fiz.}
\def\JETPE{Sov. Phys. JETP}
\def\AOP{Ann. Phys.}
\def\JPG{J. Phys. G}
\def\UJP{Ukr. J. Phys.}

\begin{document}
\title[The parametric space of the Two-Higgs-Doublet Model]
{The parametric space\\ of the Two-Higgs-Doublet Model\\
and Sakharov's baryogenesis conditions}%
\author{A.A. Kozhushko}
\affiliation{Dnipropetrovsk National University}%
\address{72, Gagarin Ave., Dnipropetrovsk 49010, Ukraine}
\email{a.kozhushko@yandex.ru; skalozubv@daad-alumni.de}
\author{V.V. Skalozub}
\affiliation{Dnipropetrovsk National University}
\address{72, Gagarin Ave., Dnipropetrovsk 49010, Ukraine}
\email{skalozubv@daad-alumni.de}

\udk{539.12.01}  \razd{\seci}

\maketitle

\begin{abstract}
The electroweak phase transition in the Two-Higgs-Doublet Model is
investigated. The Gibbs potential at finite temperature is
computed with regard for the one-loop plus ring diagram
contributions. The strong first-order phase transition satisfying
Sakharov's baryogenesis conditions is determined for the values of
scalar field masses allowed by experimental data. The relation
between the model parameters supplying the phase transition to be of
the first order is derived. It is shown that a sequence of phase
transitions is also possible. A comparison with results of other
authors is done.
\end{abstract}

%
%
\section{Introduction}
\label{intro} Nowadays it is well known that the Minimal Standard
Model (MSM) of elementary particles cannot explain the baryon
asymmetry observed in the Universe. The mechanism of generation of
this asymmetry from the initially symmetric state was proposed by
A.D. Sakharov (\cite{Sakharov}, see also review
~\cite{RubakovShaposhnikov}) and today is formulated as three
baryogenesis conditions:

1. Baryon number non-conservation.

2. C- and CP-symmetry violation.

3. Deviation from thermal equilibrium.

In electroweak theory a deviation from the thermal equilibrium can be
provided by the electroweak phase transition (EPT). The
investigations~\cite{Carrington,Shaposhnikov_ea,Demchik_Skalozub}
showed that the EPT in the MSM is strong enough for the Higgs boson
mass values that are incompatible with the modern experimental bound
$m_h \geq 114.4$ GeV. Monte Carlo simulations were used to study
the EPT in the effective three-dimensional gauge theory~\cite{Shaposhnikov_ea}.
It was shown that a critical point exists in this model. For the
experimentally allowed $m_h$ values the EPT becomes of the second order.
In~\cite{Demchik_Skalozub} the EPT in the MSM in presence of external
magnetic and hypermagnetic fields was investigated. It was concluded
that the third baryogenesis condition is not fulfilled, and the EPT
becomes of the second order in strong fields. Thus, a deviation from
the thermal equilibrium in the MSM is not strong enough, and Sakharov's
baryogenesis scenario is not realized.

In connection with this, the investigation of the EPT in  extensions
of the Standard Model is of substantial interest. One of the
extensions is the Two-Higgs-Doublet Model
(THDM)~\cite{Lee,Gunion_ea}. The THDM predicts four additional
scalar particles: a neutral particle $H$, a pair of charged fields
$H^\pm,$ and a pseudoscalar particle $A_0$. As compared with the
MSM, the THDM contains more free parameters in the scalar sector.

The EPT in the THDM was investigated in~\cite{TurokZadrozny}.
The obtained results were revised in~\cite{ClineLemieux} by using an
improved approximation (avoidance of the high-temperature expansion,
inclusion of the ring diagrams, and taking the experimentally
measured mass of the top quark into account). It was concluded that the third
baryogenesis condition is fulfilled in the THDM for some parameter values.
In \cite{Kanemura_ea_2004} a connection between the strength of the EPT and the
one-loop correction to the triple self-coupling of the lightest Higgs boson
was considered, and the deviation of the coupling value from the SM prediction was estimated.

In \cite{Land1992,Hammerschmitt_ea_1994}, as well as in the recent
paper~\cite{GinIvKan}, possible scenarios of the model behavior
during the cooling were investigated. It was shown that the realization
of a certain scenario strongly depends on the parameter values of the
scalar potential. The authors have computed the Gibbs potential in
the THDM in a simple approximation considering only the first non-trivial
finite-temperature corrections to the tree-level potential. Within this
potential they observed a few interesting phenomena:

1. A sequence of phase transitions (a second-order EPT breaks the
electroweak symmetry; then a first-order EPT occurs).

2. Exotic CP-breaking or charge-breaking minima can be realized
during the cooling.

The main goal of the present paper is to determine the domain of the
THDM parameters, for which the strong first-order phase transition happens
and the third baryogenesis condition holds. We use the approximation
of thermal equilibrium. Following \cite{Carrington,Demchik_Skalozub,ClineLemieux},
we compute the Gibbs potential in the one-loop order and include
contributions of the ring diagrams. This consistent approximation,
ensuring the minima of the effective potential to be real, allows us
to check the results of \cite{GinIvKan}.

We answer the question which is the domain in the space of model parameters
that ensures the strong first-order phase transition. To estimate that, we
introduce a certain relation between the model parameters and show that if
this relation is satisfied, the system undergoes a strong first-order phase
transition.

In \cite{TurokZadrozny,ClineLemieux,Kanemura_ea_2004} the special restrictions 
on the THDM parameters were imposed, and the authors assumed that these 
restrictions ensure the symmetry breaking along the $\tan \beta = v_2/v_1 = 1$
direction. Here, $v_i$ are the vacuum expectation values of the doublets. 
This was done to avoid difficulties with the two-stage phase
transition \cite{Land1992,Hammerschmitt_ea_1994}. We do not
restrict our investigation to this specific case. Below we show that the
noted assumption is not true in general, and a sequence of phase transitions
may occur. The phase transitions happening may be either of the first-
or second-order. In any case, the jump of the order parameter may differ
essentially from that observed in \cite{ClineLemieux}.

The paper is organized as follows. Section \ref{sec:lagrangian} contains the
necessary information on the Lagrangian and the parametrization we use.
Section \ref{sec:Gibbs} is devoted to the computation of the Gibbs potential.
In Section \ref{sec:phase_transition_results} we present the obtained results
on the phase transition. We also discuss the results obtained in
\cite{ClineLemieux,GinIvKan} and present the relation between parameters
ensuring the first-order phase transition. Concluding remarks are given
in Section \ref{sec:conclusions}. The appendices contain a special information
used in the main text.
%
%
\section{Lagrangian}
\label{sec:lagrangian}
The THDM Lagrangian differs from the MSM one in the scalar and Yukawa sectors.
It can be written as
\[
\mathcal{L} = \mathcal{L}_{\mathrm{H}} + \mathcal{L}_{f} + \mathcal{L}_{g} + \mathcal{L}_{\mathrm{Yuk}} + \mathcal{L}_{\mathrm{gauge\,fixing+ghost}} .
\]
The scalar sector is
\begin{equation}
\mathcal L_{\mathrm{H}} = \frac{1}{2} \sum_{i=1}^{2}\left|\left({\partial}_{\mu}-\frac{ig}{2}{\sigma}_{a}{A}^{a}_{\mu}-\frac{ig\prime}{2}{Y}_{{\varphi}_{i}}{B}_{\mu}\right){\varphi}_{i}\right|^{2}-V ,
\end{equation}
where $V$ is a scalar potential. To simplify the analysis we restrict our
consideration to CP-conserving vacua only. We consider the potential
which possesses the $Z_2$ symmetry~\cite{Gunion_ea,Santos_Barroso_Diaz},
\begin{eqnarray}
\label{eq:scal_pot}
\lefteqn{V = \sum_{i=1}^{2} \left[-\frac{1}{2}{\mu}_{i}^2 {\varphi}_{i}^{\dagger} {\varphi}_{i} + {\lambda}_{i}({\varphi}_{i}^{\dagger} {\varphi}_{i} )^{2} \right]+ {\lambda}_{3} \left({\mathrm{Re}[{\varphi}_{1}^{\dagger}{\varphi}_{2}]}\right)^{2} + }\nonumber\\
\lefteqn{ + {\lambda}_{4} \left({\mathrm{Im}[{\varphi}_{1}^{\dagger} {\varphi}_{2}]}\right)^{2} + {\lambda}_{5} \left({\varphi}_{1}^{\dagger} {\varphi}_{1}\right) \left({\varphi}_{2}^{\dagger} {\varphi}_{2}\right), }\\
\lefteqn{\varphi_i = \left( \begin{array}{l} \sqrt{2} a_i^+ \\ c_i + i d_i \end{array} \right). }\nonumber
\end{eqnarray}
This means the invariance with respect to the transformation
\[
\varphi_1 \to -\varphi_1, \quad \varphi_2 \to \varphi_2.
\]

The neutral scalar fields $c_i$ acquire the non-zero vacuum
expectation values (VEV) $v_{i}$ and break the $SU(2) \times U_Y(1)$
symmetry giving masses to gauge bosons and fermions. The mass mixing
of the $a_i$, $c_i$, and $d_i$ fields takes place. The spectrum of
physical particles is obtained by the substitution
\begin{eqnarray}
\label{eq:scalar_basis}
\lefteqn{
\chi^+ = a_1^+ \cos \gamma + a_2^+ \sin \gamma, \, H^+ = -a_1^+ \sin \gamma + a_2^+ \cos \gamma ,
}\nonumber\\
\lefteqn{
h = c_1 \cos \alpha + c_2 \sin \alpha, \, H = -c_1 \sin \alpha + c_2 \cos \alpha,
}\nonumber\\
\lefteqn{
\chi_3 = d_1 \cos \delta + d_2 \sin \delta, \, A_0 = -d_1 \sin \delta + d_2 \cos \delta ,
}
\end{eqnarray}
where $\chi^\pm$ and $\chi_3$ are the Goldstone modes. The
expressions for $\alpha$, $\gamma$, and $\delta$ and their
dependence on $v_{1,2}$ are adduced in Appendix A2.

In what follows, $v_{1,2}$ denote \textit{arbitrary shifts} of
$c_{1,2}$ fields. These shifts define a ``jump'' of the order
parameter during the phase transition. The extremum points  of the
potential are denoted as $v_{0\,1,2}$. They are defined as
\begin{equation}
\left. \frac{\partial V}{\partial c_i} \right\vert_{c_i=v_{0i}} = 0.
\end{equation}
We consider the case where the VEVs are non-zero for both doublets,
i.e. $v_{0\,1,2} \neq 0$ (see Appendix A1).

The gauge-boson masses are
\[
m_{W}^2 = \frac{g^2}{4} v^2, \quad m_{Z}^2 = \frac{g^2 + g'^2}{4} v^2, \quad v^2 = v_1^2 + v_2^2.
\]

In~\cite{TurokZadrozny,ClineLemieux,GinIvKan} a similar scalar
sector is considered. The main difference is that there are the
additional terms $\mu_3^2 \varphi_1^\dagger \varphi_2 + $~h.c.
softly violating the $Z_2$ symmetry. As was noted
in~\cite{TurokZadrozny}, the influence of these terms on the EPT
strength is small (though their presence allows one to introduce an
additional CP violation). Another distinction is the parametrization
of the scalar field couplings (see Appendix~A3).

As we mentioned above, the analysis in~\cite{TurokZadrozny,ClineLemieux}
was simplified by restricting the possible tree-level parameter values to
\begin{equation}
\label{eq:CLparameters}
\mu_1^2 = \mu_2^2, \quad \lambda_1 = \lambda_2 .
\end{equation}
In what follows we consider this case separately.

The general parametrization for the Yukawa interaction is
\begin{eqnarray}
\lefteqn{
\mathcal{L}_{\mathrm{Yuk}} = -\sum_{{f}_{\mathrm{L}}}\sum_{i=1}^{2} \left\lbrace {G}_{d,i} \left[ \bar{{f}_{\mathrm{L}}}{\varphi}_{i}{({f}_{d})}_{\mathrm{R}}+{(\bar{{f}_{d}})}_{\mathrm{R}}{\varphi}_{i}^{\dagger}{f}_{\mathrm{L}} \right] \right. +
}\nonumber\\
\lefteqn{
+ \left. {G}_{u,i} \left[\bar{{f}_{\mathrm{L}}}{\varphi}_{i}^{c}{({f}_{u})}_{\mathrm{R}}+{(\bar{{f}_{u}})}_{\mathrm{R}}{\varphi}_{i}^{c\dagger}{f}_{\mathrm{L}}\right]\right\rbrace ,
}\\
\lefteqn{
f_{\mathrm{L}} = \frac{1-\gamma_5}{2} \left( \begin{array}{l}f_u \\ f_d\end{array} \right), \quad f_{\mathrm{R}}=\frac{1+\gamma_5}{2}f.
}\nonumber
\end{eqnarray}
Here, $\varphi_{i}^{c} = i\sigma_{2}\varphi^{\dagger}_i$. The heavy quarks --
$t$ and $b$ -- only are of importance for the Gibbs potential, so we neglect
the CKM mixing.

This Yukawa Lagrangian leads to the existence of flavor-changing neutral
currents (FCNC)~\cite{Gunion_ea}. According to the Glashow--Weinberg
theorem~\cite{Glashow_Weinberg}, the dangerous processes with the FCNCs
can be excluded at the tree-level if all fermions of a given electric charge
couple to no more than one Higgs doublet. The most popular parametrizations
which respect these restrictions are the THDM type I and the THDM type
II~\cite{Santos_Barroso_Diaz,Barger_ea}:
\begin{itemize}
\item in the THDM type I all fermions are decoupled from the second doublet,
i.e. $G_{d,2} = G_{u,2} = 0$;
\item in the THDM type II the $u$, $c$, and $t$ quarks couple to the first doublet,
while $d$, $s$, and $b$ couple to the second doublet, i.e. $G_{d,1} = G_{u,2} = 0$.
\end{itemize}
We consider the THDM type II parametrization. The main reason is that it
represents a low-energy limit of the Minimal Supersymmetric Standard Model.

There is an interesting feature in the THDM type II. It follows from the
expressions for the quark masses at the tree-level minimum of the potential:
\begin{equation}
m_{t,0} = G_{t,1} v_{01}, \quad m_{b,0} = G_{b,2} v_{02} .
\end{equation}
Let $v_{02}$ be one or two orders of magnitude smaller than $v_{01}$.
Then $G_{t,1}$ and $G_{b,2}$ couplings have to be of the same order of
magnitude to preserve the quark mass ratio. This is an essential difference
from the MSM and the THDM type I. By noting this, we find that in the
THDM type II case it is necessary to include the $b$ quark contributions
to the Gibbs potential and Debye masses.

To compare our results with that of~\cite{ClineLemieux} we also consider
the case of  $G_{t,1} = G_{t,2}$. In this parametrization the contribution
of the $b$ quark to the Gibbs potential is also negligibly small. We will
refer to this together with restrictions (\ref{eq:CLparameters}) as
the \textit{doublet-universal} parametrization.

In the present paper, all calculations are carried out in the Feynman--'t Hooft
gauge. The gauge-fixing functions are
\begin{eqnarray}
\label{eq:gauge_fixing_functions}
\lefteqn{
G^a = \frac{1}{\sqrt{\xi}} \left( \partial^\mu A_\mu^a + \xi \frac{ig}{4}  \sum_{i=1}^{2} \left( \varphi^{\dagger}_i \sigma^a \varphi_{0i} - \varphi_{0i}^\dagger \sigma^a \varphi_i \right) \right) ,
}\nonumber\\
\lefteqn{
G = \frac{1}{\sqrt{\xi}} \left( \partial^\mu B_\mu + \xi \frac{ig'}{4}  \sum_{i=1}^{2} \left( \varphi^{\dagger}_i  \varphi_{0i} - \varphi_{0i}^\dagger \varphi_i \right) \right) ,
}\nonumber\\
\lefteqn{
\varphi_{0i} = \left( \begin{array}{l}0 \\ v_{i} \end{array} \right).
}
\end{eqnarray}
Then, the gauge-fixing part of the Lagrangian reads
\begin{equation}
\mathcal L_{\mathrm{gauge\,fixing}} = - \frac{1}{2}\left(\sum_{a=1}^3 G^{a 2} + G^2 \right).
\end{equation}
The quadratic terms in the Faddeev--Popov sector are
\begin{eqnarray}
\lefteqn{
\mathcal L_{\mathrm{ghost}} = - \bar{u}^+ (\partial^2 + \xi m_{W}^2) u^- - \bar{u}^- (\partial^2 + \xi m_{W}^2) u^+ -
}\nonumber\\
\lefteqn{
- \bar{u}_Z (\partial^2 + \xi m_{Z}^2) u_Z - \bar{u}_A \partial^2 u_A,
}
\end{eqnarray}
where $\xi$ is a gauge-fixing parameter. For arbitrary $\xi,$ the gauge-boson
propagator is
\[
iD^{\mu\nu}(p) = - \frac{i}{p^2-m^2+i\epsilon} \left( g^{\mu\nu}+(\xi-1)\frac{p^\mu p^\nu}{p^2-\xi m^2} \right).
\]

The fermion and gauge field sectors of the Lagrangian are taken to be
\begin{eqnarray}
\lefteqn{
\mathcal{L}_{f} = i \sum_{f_{\mathrm{L}}}\bar{f}_{\mathrm{L}}\gamma^{\mu}\biggl(\partial_{\mu}-\frac{ig}{2}\sigma_{a}A_{\mu}^a-\frac{ig^{\prime}}{2}Y_{f\mathrm{L}}B_{\mu}\biggr)f_{\mathrm{L}} +
}\nonumber\\
\lefteqn{
+ i \sum_{f_{\mathrm{R}}}\bar{f}_{\mathrm{R}}\gamma^{\mu}\biggl(\partial_{\mu}-ig^{\prime}Q_{f}B_{\mu}\biggr)f_{\mathrm{R}} ,
}\nonumber\\
\lefteqn{
\mathcal{L}_{\mathrm{g}} = -\frac{1}{4}{F}^{\mu\nu}{F}_{\mu\nu}-\frac{1}{4}{F}^{\mu\nu}_a {F}_{\mu\nu}^{a} .
}
\end{eqnarray}
We proceed with the calculation of the Gibbs potential.
%
%
\section{Gibbs Potential}
\label{sec:Gibbs}
The calculation of the Gibbs potential was discussed in numerous
papers~\cite{VacGibbs,FinTempGibbs} (see also review  \cite{Sher}).
We compute the effective potential in the following standard way:
\begin{equation}
V_G(v_i)=V_{\mathrm{tree}} + V^{1}_{\mathrm{vac}}(v_i)+V^{1}_{T}(v_i),
\end{equation}
where $V_{\mathrm{tree}}$ is the tree-level potential, $V^{1}_{\mathrm{vac}}$
is the one-loop correction at zero temperature, $V^{1}_{T}$ is the one-loop
finite-temperature correction. The tree-level part is obtained by
substituting $\varphi_{0i}$ into (\ref{eq:scal_pot}) and  reads
\begin{eqnarray}
\label{eq:treelevelpotential}
\lefteqn{
V_{\mathrm{tree}}(v_i) = -\frac{1}{2}(\mu_1^2 v_1^2 + \mu_2^2 v_2^2) + \lambda_1 v_1^4 + \lambda_2 v_2^4 +
}\nonumber\\
\lefteqn{
 + (\lambda_3 + \lambda_5)v_1^2 v_2^2.
}
\end{eqnarray}

Now, we consider $V^{1}_{\mathrm{vac}}$ and $V^{1}_{T}$.
%
%
\subsection{One-loop contributions at zero temperature}
\label{subsec:ZeroTemp}
The regularized contribution of a field of mass $m$ to $V_G$
is~\cite{VacGibbs}
\begin{eqnarray}
\label{eq:vac_corr}
\lefteqn{
V^1_{\mathrm{v}}(m) = \frac{1}{64\pi^2} \biggl(\frac{m^2}{s_0} +
}\nonumber\\
\lefteqn{
+ m^4 \bigl(\ln(s_0 m^2) - 3/2 + \gamma - \frac{i\pi}{2} \bigr) \biggr).
}
\end{eqnarray}
We use Schwinger's proper time regularization with the
regularization parameter $s_0$. It has to be set to zero at the end
of calculations.

The general expression for the scalar field mass is
\begin{eqnarray}
\label{eq:general_scalar_mass}
\lefteqn{
m^2_{\pm} = B_1 v_1^2 + B_2 v_2^2 + B_3 \pm
}\nonumber\\
\lefteqn{
\pm \sqrt{(C_1 v_1^2 + C_2 v_2^2 + C_3)^2 + (D_1 v_1 v_2 )^2}.
}
\end{eqnarray}
Here, $B_i$, $C_i$, and $D_i$ are some combinations of the
tree-level VEVs and couplings. There are four pairs of the scalar
fields: $h$ and $H$, $\chi^\pm$ and $H^\pm$, $\chi_3$ and $A_0$. The
sign in Eq. (\ref{eq:general_scalar_mass}), ``$-$'' or ``$+$'',
corresponds to the mass of one field of a pair. For example, we have
the ``$-$'' sign in case of the $\chi_3$ mass and the ``$+$'' sign
for the $A_0$ mass (see Appendix A2). In the sum of both field
contributions, the term
\begin{equation}
\frac{1}{2}(m^4_{+} - m^4_{-})\ln\biggl(\frac{m^2_+}{m^2_-} \biggr)
\end{equation}
appears. It is cancelled out by the term  coming from a high-temperature
expansion, when the finite-temperature corrections are taken into account.
However, we do not use this expansion in our calculations, and, therefore,
the explicit cancellation does not occur. This term results in cumbersome
quantum corrections to $V_G$.

The contribution coming from fermions, gauge bosons and ghosts is given
by (\ref{eq:vac_corr}) with regard for the factor $A$,
\begin{equation}
\label{eq:degrees_of_freedom_multiplier}
A = \left\{
\begin{array}{ll}
-1 \times 4 \times 3, & \textrm{  quark,} \\
-2, & \textrm{  ghost,} \\
3+\xi^2, & \textrm{  gauge boson.} \\
\end{array} \right.
\end{equation}
This factor accounts for the number of degrees of freedom and the color
states of fields.

We choose the renormalization conditions preserving the tree-level
vacuum energy value, VEVs, and mass terms. They are taken to be
\begin{eqnarray}
\lefteqn{
V_{\mathrm{v}}(v_{0i}) = V_{\mathrm{tree}}(v_{0i}), \qquad  \frac{\partial V_{\mathrm{v}}}{\partial v_{i}} \biggr|_{\mathrm{vac}} = 0,
}\nonumber\\
\lefteqn{
\frac{\partial^2 V_{\mathrm{v}}}{\partial v_{1}^2} \biggr|_{\mathrm{vac}} = -\mu_1^2 + 12 \lambda_1 v_{01}^2 + 2 (\lambda_3 + \lambda_5) v_{02}^2,
}\nonumber\\
\lefteqn{
\frac{\partial^2 V_{\mathrm{v}}}{\partial v_{2}^2} \biggr|_{\mathrm{vac}} = -\mu_2^2 + 12 \lambda_2 v_{02}^2 + 2 (\lambda_3 + \lambda_5) v_{01}^2,
}\nonumber\\
\lefteqn{
\frac{\partial^2 V_{\mathrm{v}}}{\partial v_{1} \partial v_{2}} \biggr|_{\mathrm{vac}} = 0.
}
\end{eqnarray}

Since the renormalized contributions from the scalar sector are cumbersome,
we do not adduce them here. They could be obtained easily by using
a symbolic calculation software. The renormalized contributions of a fermion,
a gauge boson, or a ghost field read
\begin{eqnarray}
\lefteqn{
V^{1,\, \mathrm{r}}_{\mathrm{v}} (m) = \frac{A}{64\pi^2} \biggl(m^4 \bigl(\ln(\frac{m^2}{m^2_{\mathrm{vac}}}) - \frac{1}{2} \bigr) + \frac{m^4_{\mathrm{vac}}}{2} -
}\nonumber\\
\lefteqn{
- (m^2 - m^2_{\mathrm{vac}})^2 \biggr),
}
\end{eqnarray}
where $m_{\mathrm{vac}}$ is the field mass value at $v_i=v_{0i}$.

The complete temperature-independent part of the Gibbs potential is
\begin{eqnarray}
\label{eq:vacuum_potential_full}
\lefteqn{
V_{\mathrm{v}}(v_i) = V_{\mathrm{tree}} + V_{\mathrm{v},\,h,H}^{1,\, \mathrm{r}} + V_{\mathrm{v},\,\chi^{\pm},H^{\pm}}^{1,\, \mathrm{r}} + V_{\mathrm{v},\,\chi_3,A_0}^{1,\, \mathrm{r}} +
}\nonumber\\[2mm]
\lefteqn{
+ 2(3+\xi^2)V_{\mathrm{v}}^{1,\, \mathrm{r}}(m_W) + (3+\xi^2)V_{\mathrm{v}}^{1,\, \mathrm{r}}(m_Z) -
}\nonumber\\[2mm]
\lefteqn{ - 4 V_{\mathrm{v}}^{1,\, \mathrm{r}}(\sqrt{\xi} m_W) - 2
V_{\mathrm{v}}^{1,\, \mathrm{r}}(\sqrt{\xi} m_Z) -
}\nonumber\\[2mm]
\lefteqn{ - 12 V_{\mathrm{v}}^{1,\, \mathrm{r}}(m_t) - 12
V_{\mathrm{v}}^{1,\, \mathrm{r}}(m_b), }
\end{eqnarray}
where $V_{\mathrm{v},\,h,H}^{1,\, \mathrm{r}}$, 
$V_{\mathrm{v},\,\chi^{\pm},H^{\pm}}^{1,\, \mathrm{r}}$ and 
$V_{\mathrm{v},\,\chi_3,A_0}^{1,\, \mathrm{r}}$ are the scalar field 
contributions.
%
%
\subsection{One-loop contributions at finite temperatures}
\label{subsec:temp_corr}

Finite-temperature corrections are calculated by using the Matsubara
formalism. For the contribution of one bosonic degree of freedom we
have~\cite{FinTempGibbs}
\begin{equation}
\label{eq:boson_temp_corr} V^{\mathrm{b}}_{T}(m) =
\frac{T^4}{2\pi^2}\int\limits_0^{\infty} dx\,x^2 \ln
\left(1-\exp(-\sqrt{x^2+\beta^2 m^2}) \right),
\end{equation}
where $\beta$ is the inverse temperature. We recall that Matsubara's frequencies
for ghost fields are even \cite{LandsmanWeert}.
For one fermionic degree of freedom, we have
\begin{equation}
\label{eq:fermion_temp_corr} V^{\mathrm{f}}_{T}(m) =
\frac{T^4}{2\pi^2}\int\limits_0^{\infty} dx x^2 \ln
\left(1+\exp(-\sqrt{x^2+\beta^2 m^2}) \right).
\end{equation}
The degree-of-freedom factors for fermions and ghosts stand in
(\ref{eq:degrees_of_freedom_multiplier}). The contribution of
the massive gauge field is
\begin{equation}
\label{eq:vector_boson_temp_corr}
V^{\mathrm{gauge}}_{T}(m) = 3 V^{\mathrm{b}}_{T}(m) + V^{\mathrm{b}}_{T}(\sqrt{\xi}m).
\end{equation}
The last term in (\ref{eq:vector_boson_temp_corr}) cancels a part of
the ghost field contribution.
%
%
\subsection{Ring diagram contributions}
\label{subsec:ring}

\begin{figure}
\includegraphics[width=\column]{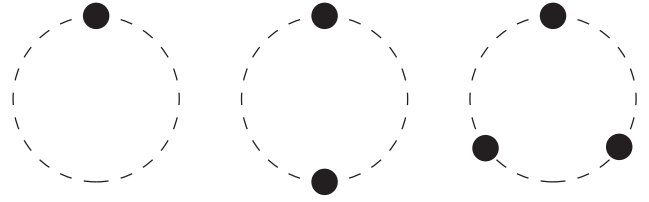}
\vskip-3mm\caption{Ring-diagram contributions for scalar field.
Black blobs denote Debye masses} \label{fig:Daisy}
\end{figure}

As is well known, an imaginary part of the one-loop Gibbs
potential arises at small $v_i$. It comes from the scalar sector
contributions at finite temperature and indicates the instability
of the system. Gauge bosons are massless in the symmetric phase
that leads to infrared divergences. These shortcomings of the
one-loop effective potential can be avoided by adding the
ring-diagram (Fig. \ref{fig:Daisy}) contributions
\cite{FinTempGibbs,Fradkin,KirzhnitsLinde,Kapusta}.
These diagrams introduce additional finite-temperature corrections
to the masses of bosons that result in the terms of the order
$\sim g^3$ or $\sim \lambda_i^{3/2}$ in $V_G$.

The ring-improved finite-temperature correction to the Gibbs potential
in the scalar field case is
\begin{eqnarray}
\lefteqn{ V^{\mathrm{b}}_{T}(m(T))\! =\!
\frac{T^4}{2\pi^2}\!\int\limits_0^{\infty}\! dx\,x^2 \ln
\left(1\!-\!\exp(-\sqrt{x^2\!+\!\beta^2 m^2(T)})\right),
}\nonumber\\
\lefteqn{
m^2(T) = m^2 + \delta m^2(T),
}
\end{eqnarray}
where $\delta m^2(T)$ denotes the Debye mass of a field. This
correction ensures that the imaginary part for the finite-temperature
contribution is absent. The Debye mass is defined through the
polarization tensor $\Pi$ of a field taken in the infrared
limit \cite{Kapusta}
\begin{equation}
\delta m^2(T) = \Pi(k_0 = 0, \bar{k} \to 0),
\end{equation}
where $k$ is the four-momentum of a field.

For gauge fields, the Debye mass is defined as $-\Pi_{00}$ in
the infrared limit. The ring-diagram contribution to $V_G$
from each massive gauge boson is \cite{Carrington}
\begin{equation}
V^{\mathrm{ring}}_{\mathrm{g}}(m) = -\frac{T}{12\pi}\left((m^2 + \delta m^2(T))^{3/2}-m^3 \right).
\end{equation}

The one-loop Debye masses of the Higgs fields in the THDM type II are
\begin{eqnarray}
\label{eq:debye_masses_h_H}
\lefteqn{
\delta m^2_{h}(T) = T^2 \left((2\lambda_1 + \frac{1}{2} G_t^2) \cos^2 \alpha \right. + (2\lambda_2 +
}\nonumber\\
\lefteqn{
+ \frac{1}{2} G_b^2) \sin^2 \alpha + \left. \frac{\lambda_3 + \lambda_4 + 4\lambda_5}{6} + \frac{3g^2 + g'^2}{16} \right),
}\nonumber\\
\lefteqn{
\delta m^2_{H}(T) = T^2 \left((2\lambda_1 + \frac{1}{2} G_t^2) \sin^2 \alpha \right. + (2\lambda_2 +
}\nonumber\\
\lefteqn{ + \frac{1}{2} G_b^2) \cos^2 \alpha + \left.
\frac{\lambda_3 + \lambda_4 + 4\lambda_5}{6} + \frac{3g^2 +
g'^2}{16} \right). }
\end{eqnarray}
For other scalar fields, the Debye masses are given by similar expressions.
The difference is that the angle $\alpha$ is replaced by $\gamma$, $\delta$
from Eq. (\ref{eq:scalar_basis}).

The complete Higgs-sector contribution to the finite-temperature part
of $V_G$ is
\begin{eqnarray}
\lefteqn{
V^{\mathrm{s}}_T(v_i) = V^{\mathrm{b}}_{T}(m_{h}(T)) + V^{\mathrm{b}}_{T}(m_H(T)) + 2 V^{\mathrm{b}}_{T}(m_{\chi^{\pm}}(T))+
}\nonumber\\
\lefteqn{
+ 2 V^{\mathrm{b}}_{T}(m_{H^{\pm}}(T)) + V^{\mathrm{b}}_{T}(m_{\chi_3}(T)) + V^{\mathrm{b}}_{T}(m_{A_0}(T)).
}
\end{eqnarray}
The gauge-boson Debye masses are
\begin{equation}
\label{eq:gauge_debye}
\delta m^2_{W}(T) = 2 g^2 T^2, \quad \delta m^2_{Z}(T) = \frac{11 g'^4 + 5 g^4}{4(g^2 + g'^2)}T^2.
\end{equation}

For Faddeev--Popov ghosts, the Debye mass is zero in the leading
order in $T$.

In  Appendix B we give the Debye masses of all fields. We note that
these corrections are $\xi$-independent.

The finite-temperature gauge-field and ghost contribution to $V_G$ is
\begin{eqnarray}
\lefteqn{
V^{\mathrm{g}}_T(v_i) = 6 V^{\mathrm{b}}_{T}(m_W) + 3V^{\mathrm{b}}_{T}(m_Z) - 2 V^{\mathrm{b}}_{T}(\sqrt{\xi} m_W) -
}\nonumber\\
\lefteqn{
- V^{\mathrm{b}}_{T}(\sqrt{\xi} m_Z) + 2 V^{\mathrm{ring}}_{\mathrm{g}}(m_W) + V^{\mathrm{ring}}_{\mathrm{g}}(m_Z).
}
\end{eqnarray}
For the final expression of $V_G$ in the THDM we have
\begin{equation}
V_G(v_i) = V_{\mathrm{v}}(v_i) + V^{\mathrm{s}}_T(v_i) + V^{\mathrm{g}}_T(v_i) - 12 V^{\mathrm{f}}_{T}(m_t) - 12 V^{\mathrm{f}}_{T}(m_b).
\end{equation}

\begin{figure}
\includegraphics[width=\column]{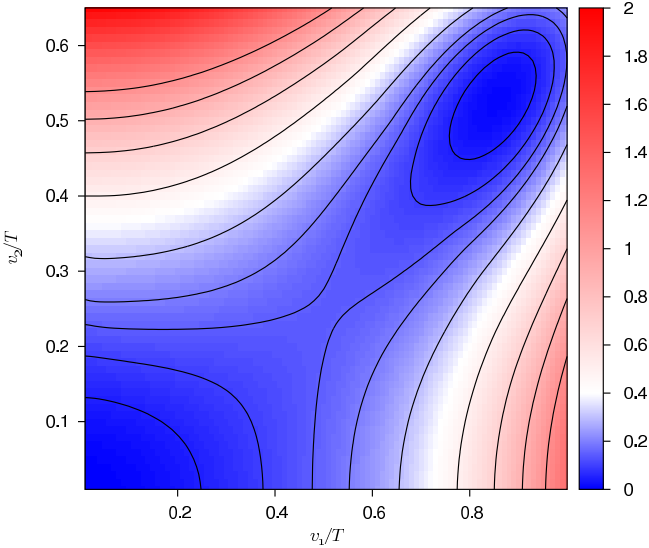}
\vskip-3mm\caption{Two distinct minima signaling  the EPT  of the
first order are realized. The critical temperature is $T_c =
125.65~\mathrm{GeV}$} \label{fig:1st_order_no_opt}
\end{figure}

The minimum value of $V_G$ is gauge-independent. The $\xi$-dependence
is cancelled out between the gauge boson, Goldstone field and ghost
contributions. Physical quantities are also
gauge-invariant \cite{GaugeDependence}.

However, the Gibbs potential is gauge-dependent \cite{VacGibbs} at
arbitrary values of $v_i$. For numerical calculations one has to choose
the value for $\xi$. We set $\xi=1$. In this case the Goldstone field
masses equal to the masses of corresponding gauge bosons.
%
%
\section{Phase Transition}
\label{sec:phase_transition_results}
The third Sakharov condition is fulfilled if the first-order EPT is
realized, and the order parameter jump is greater than
1 \cite{RubakovShaposhnikov}:
\begin{equation}
\frac{\delta v}{T_c} > 1,
\end{equation}
where $T_c$ is the critical temperature.

In the THDM there are two order parameters -- $v_1/T$ and $v_2/T$.
In general, several jumps of the order parameters may occur, i.e.
single phase transitions, as well as sequences of phase transitions,
may happen. During a series of EPTs the system goes to an intermediate
vacuum state, where the symmetry is broken for one doublet only. We
discuss possible scenarios of phase transitions in the THDM and find
the domain in the parameter space, for which $\delta v / T_c$ is large.

We consider $\lambda_{1,2,3,4,5}$ and one of the VEVs $v_{01,2}$ as
the free parameters of the model (see Appendix A1).
%
%
\subsection{Possible scenarios}
The plots of  $100(V_G(v_1,v_2)-V_G(0,0))/T^4$ versus $v_{1}/T$,
$v_{2}/T$ are shown in the figures. The blue areas represent lower
values of the Gibbs potential. The parameter values for the figures
can be found in Appendix~C.

First, we consider the THDM type II.

1. Fig. \ref{fig:1st_order_no_opt}. The form of $V_G$ indicates
a strong first-order EPT. The order parameter jump is
\[
\delta v = \frac{\sqrt{\delta v_1^2 + \delta v_2^2}}{T_c} = 1.01.
\]
For this set of parameter values, the tree-level scalar field masses
are
\begin{eqnarray}
\lefteqn{ m_h = 119~\mathrm{GeV}, \quad m_H = 131~\mathrm{GeV},
\quad m_{H^{\pm}} = 181~\mathrm{GeV},
}\nonumber\\
\lefteqn{ m_{A_0} = 338~\mathrm{GeV}. }\nonumber
\end{eqnarray}
This scenario is the most favorable for successful baryogenesis.

\begin{figure*}
\begin{minipage}{0.49\textwidth}
  \includegraphics[width=\column]{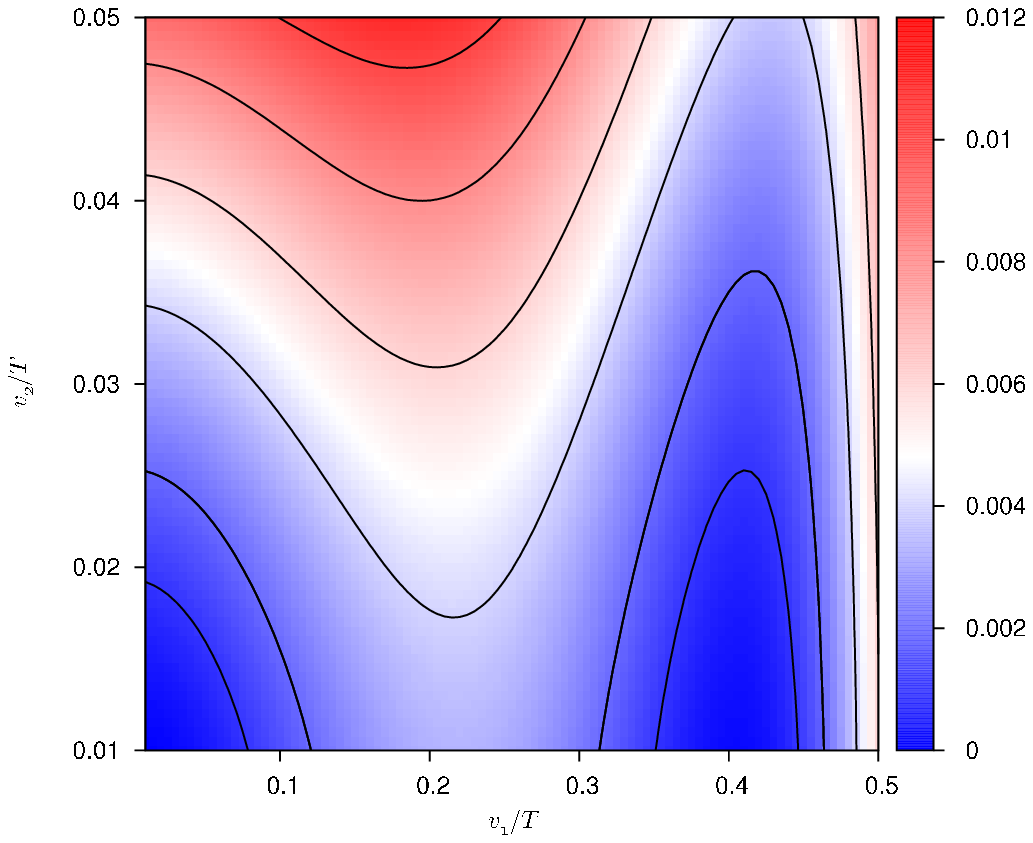}\\ \centering{\large{\it a}}
\end{minipage}
\hfill
\begin{minipage}{0.49\textwidth}
  \includegraphics[width=\column]{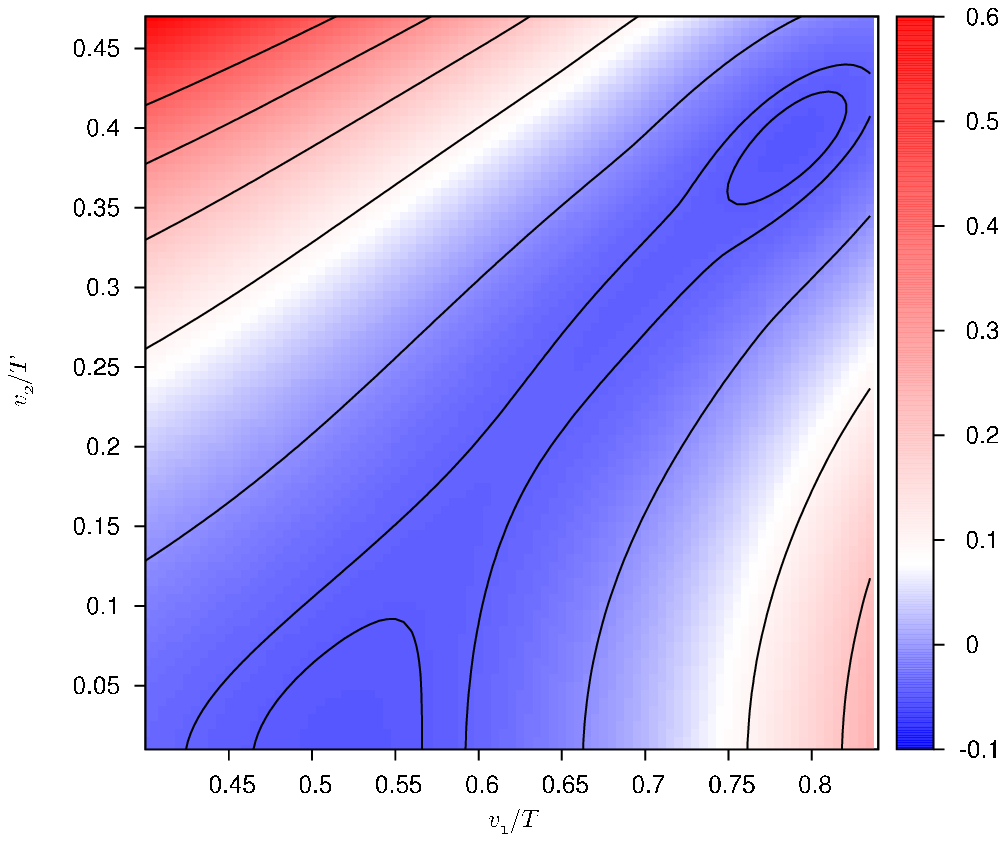}
\\ \centering{\large{\it b}}
\end{minipage}
\vskip-2mm \caption{Sequence of the first-order phase transitions.
The critical temperatures are ({\it a}) $T_c = 128.18~\mathrm{GeV}$,
({\it b}) $T_c = 127.66~\mathrm{GeV}$}\vskip3mm
\label{fig:1st_order_sequence}
\end{figure*}

2. Fig.~\ref{fig:1st_order_sequence}. The sequence of phase
transitions is generated. The weak first-order EPT breaking  the
symmetry along the directions $\tan \beta = 0$ or $\tan \beta =
+\infty$ happens (the former case is shown in Fig. 3,{\it a}). Then
next weak first-order EPT follows (Fig. 3,{\it b}). The Gibbs
potential minimum is now located along the $0 < \tan \beta <
+\infty$ direction in the $(v_1 ,\,v_2)$ plane. The scalar field
masses in the tree-level approximation are
\begin{eqnarray}
\lefteqn{ m_h = 114~\mathrm{GeV}, \quad m_H = 132~\mathrm{GeV},
\quad m_{H^{\pm}} = 181~\mathrm{GeV},
}\nonumber\\
\lefteqn{ m_{A_0} = 266~\mathrm{GeV} . }\nonumber
\end{eqnarray}
The successful baryogenesis cannot be realized. The thermal equilibrium
approximation used could be unreliable. This is because of non-equilibrium
processes happening after the first phase transition. In fact, this
series of phase transitions, as concerns its consequences, could substitute
one strong enough first-order phase transition. The calculation of characteristics
for a phase transition of such type requires other methods and additional
investigations.

\begin{figure}
\includegraphics[width=\column]{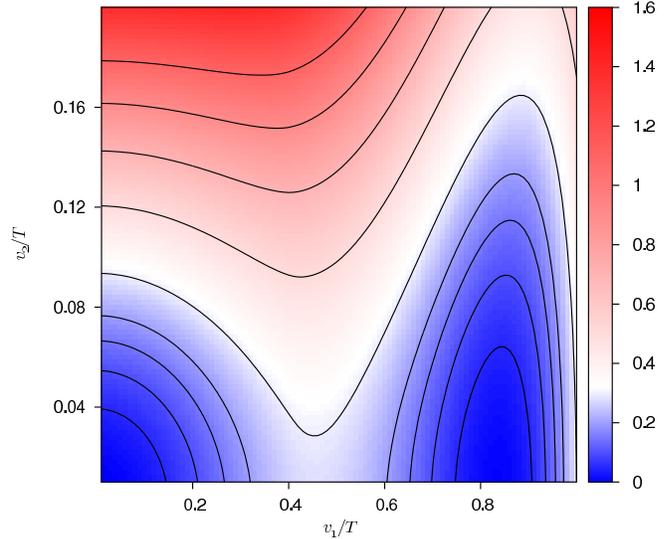}
\vskip-3mm\caption{The first-order EPT in the sequence of phase
transitions. The critical temperature is $T_c = 153.3~\mathrm{GeV}$}
\label{fig:1st2nd_order_sequence_1}
\end{figure}

3. Fig. \ref{fig:1st2nd_order_sequence_1}. The sequence of phase
transitions happens. A strong first-order EPT breaking of the symmetry
along the $\tan \beta = 0$ or $\tan \beta = +\infty$ direction is
generated (the former case is shown in the figure). Then the system
undergoes a second-order phase transition, the $\tan \beta $ value
being finite and non-zero. The scalar field masses are
\begin{eqnarray}
\lefteqn{ m_h = 120~\mathrm{GeV}, \quad m_H = 201~\mathrm{GeV},
\quad m_{H^{\pm}} = 322~\mathrm{GeV},
}\nonumber\\
\lefteqn{ m_{A_0} = 429~\mathrm{GeV} . }\nonumber
\end{eqnarray}
This scenario is acceptable for the successful baryogenesis.

4. A sequence of second-order phase transitions is possible. Such
scenario is realized, for example, if the scalar field masses at
the tree level are
\begin{eqnarray}
\lefteqn{ m_h = 114~\mathrm{GeV}, \quad m_H = 162~\mathrm{GeV},
\quad m_{H^{\pm}} = 181~\mathrm{GeV},
}\nonumber\\
\lefteqn{ m_{A_0} = 276~\mathrm{GeV} . }\nonumber
\end{eqnarray}
In this case the baryogenesis is not realized.

These results also hold for the THDM type I qualitatively and, in most
cases, quantitatively. The main difference from the THDM type II is that
the $b$ quark influence on the Gibbs potential is much weaker. Therefore,
a quantitatively different picture should be expected for small
$\tan \beta_0 = v_{02}/v_{01}$ values, namely $\tan \beta_0 \leq 0.1$.
For this domain of the parameter space the $b$ quark contribution to $V_G$
in the THDM type II is non-negligible.

Note that the parameter values considered correspond to Sector I of
the parameter space defined in \cite{GinIvKan}. In this sector, the authors
observed a second-order EPT. They proposed that a more detailed
investigation based on a consistent corrected effective potential could
predict the EPT of the first order. This is just what we have observed.

\begin{figure*}
\begin{minipage}{0.49\textwidth}
  \includegraphics[width=\column]{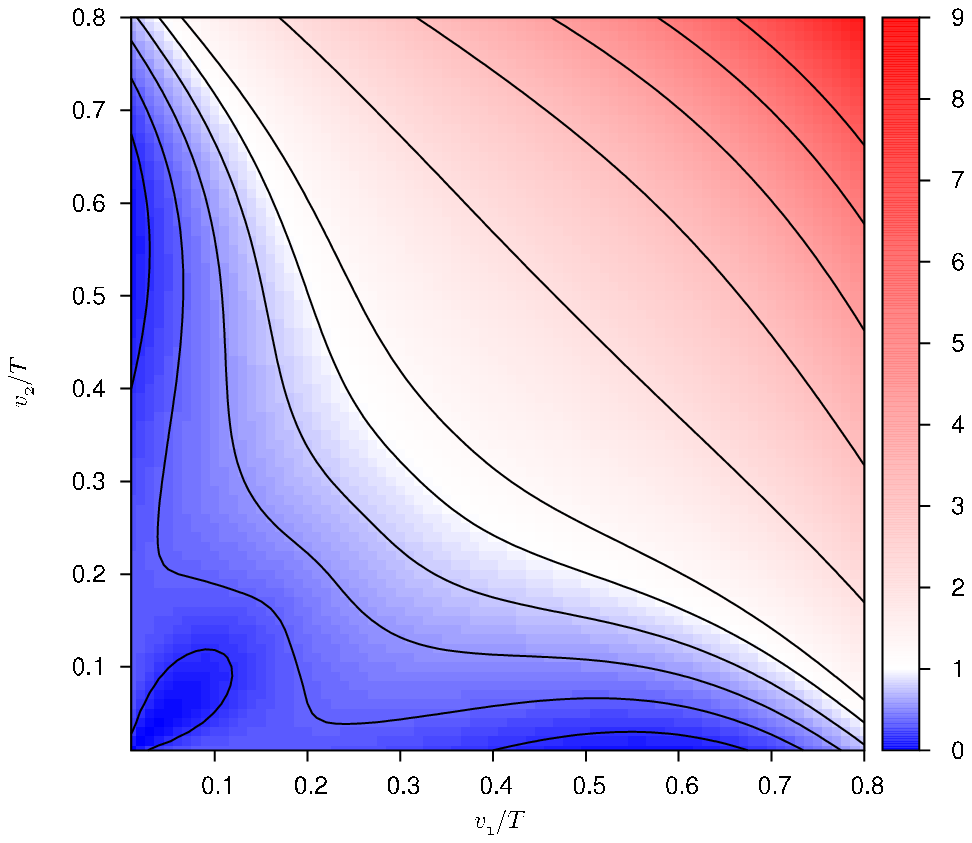}
\\ \centering{\large{\it a}}
\end{minipage}
\hfill
\begin{minipage}{0.49\textwidth}
  \includegraphics[width=\column]{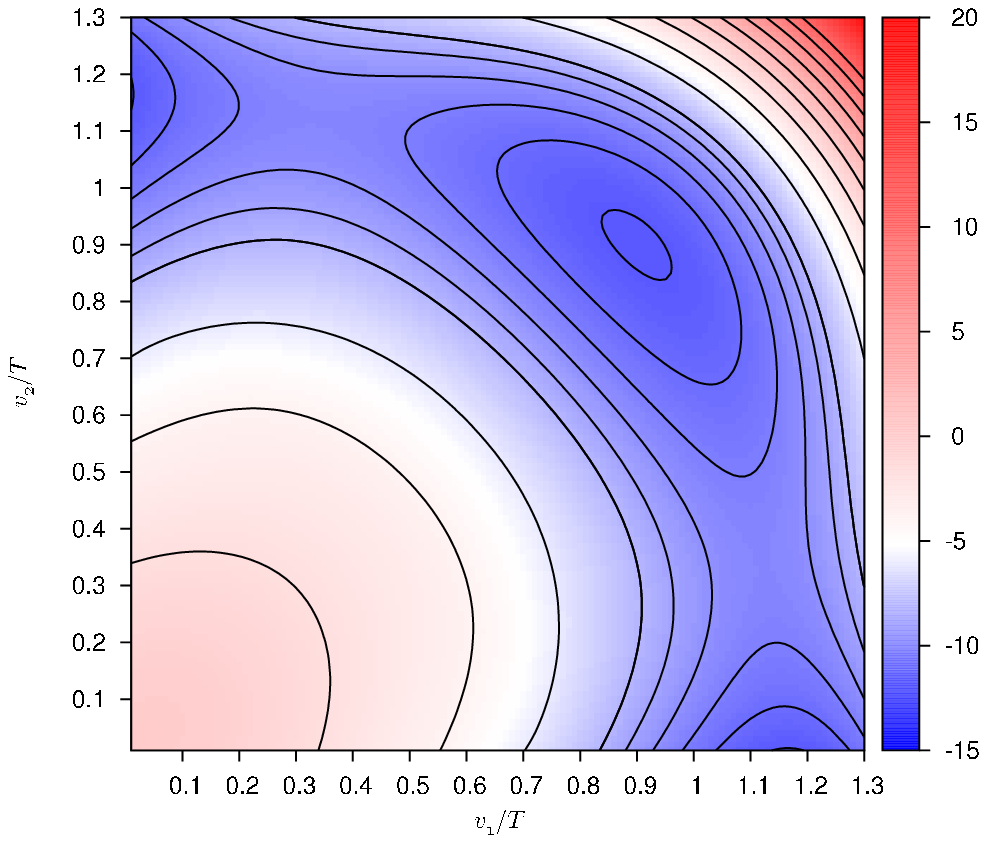}
\\ \centering{\large{\it b}}
\end{minipage}
\vskip-2mm\caption{Sequence of the first-order EPTs  is realized.
The critical temperatures are ({\it a}) $T_c = 192~\mathrm{GeV}$;
({\it b}) $T_c = 154~\mathrm{GeV}$}
\label{fig:1st_order_sequence_Cline}
\end{figure*}
%
%
\subsection{The doublet-universal parametrization}
Since there is a possibility of a sequence of phase transitions, it is
interesting to discuss the assumption about the symmetry breaking
made in \cite{TurokZadrozny,ClineLemieux,Kanemura_ea_2004}. In \cite{ClineLemieux}, 
a strong first-order EPT was observed. It was assumed that the symmetry
breaking happens along the $\tan \beta = 1$ direction, if restrictions
(\ref{eq:CLparameters}) are imposed. This allows one to express all free
parameters of the potential in terms of the tree-level masses of fields.
The domain of the model parameter space considered in \cite{ClineLemieux}
corresponds to the condition for scalar field masses
\begin{equation}
\label{eq:CL_mass_relation}
m_H = m_{H^\pm} = m_{A_0}.
\end{equation}

In \cite{ClineLemieux} the tree-level masses of scalar fields were
used as free parameters instead of the couplings. However, the mass
values do not define uniquely the  values of couplings. Really, let
us consider the $h$ field mass case. The tree-level expression of
$m_h$ at the minimum point of the potential is
\begin{eqnarray}
\lefteqn{
m_h^2 = 4 \lambda_1 v_{01}^2 + 4 \lambda_2 v_{02}^2 -
}\nonumber\\
\lefteqn{
- 4\sqrt{( \lambda_2 v_{02}^2 -  \lambda_1 v_{01}^2)^2 + (v_{01} v_{02} (\lambda_3 + \lambda_5))^2}.
}\nonumber
\end{eqnarray}
By applying (\ref{eq:CLparameters}), we obtain
\begin{equation}
m_h^2 = 4 v_{01}^2 \left(2 \lambda_1 - | \lambda_3 + \lambda_5 | \right).
\end{equation}
Since $\lambda_3 < 0$ (see Appendix A2), there exist two possible
values of $\lambda_5$ corresponding to the same value of $m_h$,
namely
\begin{equation}
\lambda_5^\pm = - \lambda_3 \pm (2 \lambda_1 - \frac{m_h^2}{4 v_{01}^2}).
\end{equation}
It appears that, at $\lambda_5 = \mathrm{Min} (\lambda_5^+, \, \lambda_5^-)$
for the parameter values resulting in (\ref{eq:CL_mass_relation}), the symmetry
is broken along the $\tan \beta = 1$ direction, as it was assumed. However,
if $\lambda_5$ is taken to be $\mathrm{Max} (\lambda_5^+, \, \lambda_5^-)$,
then the evolution of the system is completely different. The system undergoes
the sequence of first-order phase transitions. The  minima with broken
symmetry appear along the $\tan \beta = 0$ and $\tan \beta = +\infty$ directions.
After that, the Gibbs potential develops another minimum along the $\tan \beta = 1$
direction. This sequence is shown in Fig. \ref{fig:1st_order_sequence_Cline}.
We can conclude that the assumption mentioned is not always true.
These model parameters correspond to the scalar field mass values
\[
m_h = 120~\mathrm{GeV}, \quad m_H = m_{H^{\pm}} = m_{A_0} =
250~\mathrm{GeV}.
\]

Note also that in this parametrization three degenerate vacuum states
may coexist.

%
%
\subsection{Relation between model parameters}
\label{subsec:toymodel} It can be seen from Appendix C that a small
change of any parameter value may result in a significant change  in
the system's evolution (compare the scenarios shown in Figs.
\ref{fig:1st_order_no_opt} and \ref{fig:1st_order_sequence}). So it
is interesting to determine any relation between model parameters
which provides a large $\delta v/T_c$ value, and, hence, a strong
first-order phase transition. Let us turn to this problem.

From  geometric reasons it is natural to assume that a jump of the
order parameter is large enough if the nontrivial vacuum appears in the
$0 < \tan \beta < +\infty$ direction. This happens if the symmetry is
broken for both Higgs doublets simultaneously. The analytical form of
this condition reads
\begin{equation}
\label{eq:opt_condition}
\left\{ \begin{array}{l}
\mu_1^2(T) = 0, \\ \mu_2^2(T) = 0.
\end{array} \right.
\end{equation}
The values of couplings ensuring this condition can be found by using
the high-temperature expansion of~(\ref{eq:boson_temp_corr}) and
(\ref{eq:fermion_temp_corr}).

Since the quantum corrections are cumbersome in the THDM, we discuss the
application of (\ref{eq:opt_condition}) considering a toy model as an
example. Then we apply this relation to the THDM type II.

The toy model has to possess two main properties of the THDM -- the
spontaneous symmetry breaking and the mass mixing of scalar fields. The
Lagrangian we use is
\begin{eqnarray}
\lefteqn{
\mathcal{L} = \frac{1}{2}\partial_{\mu} \varphi_1 \partial^{\mu} \varphi_1 + \frac{1}{2}\partial_{\mu} \varphi_2 \partial^{\mu} \varphi_2 - V,
}\nonumber\\
\lefteqn{
V = -\frac{1}{2}(\mu_1^2 \varphi_1^2 + \mu_2^2 \varphi_2^2) + \lambda_1 \varphi_1^4 + \lambda_2 \varphi_2^4 + \lambda_3 \varphi_1^2 \varphi_2^2,
}
\end{eqnarray}
where $\varphi_i$ are real scalar fields. The Lagrangian is invariant with
respect to the transformation $\varphi_i \longrightarrow -\varphi_i$.

Let us shift $\varphi_i$ by arbitrary values $v_i$ and obtain the mass
eigenstates as those in Eq. (\ref{eq:scalar_basis}). We denote a new pair of fields
as $h_{1,2}$. The tree-level VEVs $v_{0i}$ are obtained from the relations
\begin{equation}
\left\lbrace \begin{array}{l}
-\mu_1^2 + 4\lambda_1 v_1^2 + 2 \lambda_3 v_2^2 = 0, \\
-\mu_2^2 + 4\lambda_2 v_2^2 + 2 \lambda_3 v_1^2 = 0 . \nonumber
\end{array} \right.
\end{equation}
The tree-level potential is
\begin{equation}
V_{\mathrm{tree}} = -\frac{1}{2}(\mu_1^2 v_1^2 + \mu_2^2 v_2^2) + \lambda_1 v_1^4 + \lambda_2 v_2^4 + \lambda_3 v_1^2 v_2^2 .
\end{equation}
The one-loop corrections to the Gibbs potential are given by Eqs.
(\ref{eq:vac_corr}) and (\ref{eq:boson_temp_corr}). We use the
$\overline{MS}$ renormalization scheme. Then the one-loop vacuum
corrections can be written as
\[
V^1_{\mathrm{vac}}(m) = \frac{1}{64 \pi^2} m^4 \ln \left( \frac{m^2}{\mu^2} \right) .
\]
The high-temperature expansion of (\ref{eq:boson_temp_corr}) looks
as follows:
\begin{eqnarray}
\lefteqn{
V^{\mathrm{b}}_T (m) = -\frac{\pi^2 T^4}{90} + \frac{T^2 m^2}{24} - \frac{T m^3}{12 \pi} -
}\nonumber\\
\lefteqn{
- \frac{1}{64 \pi^2} m^4 (\ln \frac{m^2}{T^2} - 5.41) .
}\nonumber
\end{eqnarray}
For the one-loop Gibbs potential, we have
\[
V_G^{\mathrm{toy}} = V^1_{\mathrm{vac}}(m_{h_1}) +
V^1_{\mathrm{vac}}(m_{h_2}) +
\]
\begin{equation}
\label{eq:toy_gibbs} +V^{\mathrm{b}}_T(m_{h_1}) +
V^{\mathrm{b}}_T(m_{h_2}).
\end{equation}
The functions $\mu_{1,2}(T)$ are the factors at $v_{1,2}^2$ in (\ref{eq:toy_gibbs}):
\begin{eqnarray}
\lefteqn{
\mu_{1}^2(T) = \mu_1^2 + \frac{1}{64\pi^2} (12\lambda_2 \mu_1^2 + 2 \lambda_3 \mu_2^2) \times
}\nonumber\\
\lefteqn{
\times \left(\ln\frac{T^4}{\mu^4}+10.82 \right) - \frac{T^2}{12}(12 \lambda_1 + 2 \lambda_3),
}\nonumber\\
\lefteqn{
\mu_{2}^2(T) = \mu_2^2 + \frac{1}{64\pi^2} (12\lambda_1 \mu_2^2 + 2 \lambda_3 \mu_1^2) \times
}\nonumber\\
\lefteqn{
\times \left(\ln\frac{T^4}{\mu^4}+10.82 \right)  - \frac{T^2}{12}(12 \lambda_2 + 2 \lambda_3).
}
\end{eqnarray}

There are seven parameters in the toy model -- $\lambda_{1,2,3}$,
$\mu_{1,2}$, $v_{01,2}$. Four of them are free. Note that we took
the condition $v_{01}^2 + v_{02}^2 = \mathrm{const}$ into account.
The way of using (\ref{eq:opt_condition}) is the following. We take
arbitrary numerical values for any three parameters (for example,
$\lambda_{1,2,3}$), and the value of the remaining parameter ($v_{01}$)
is determined by solving the system (\ref{eq:opt_condition}).

The fulfillment of (\ref{eq:opt_condition}) is sufficient for the
symmetry to be broken along the $0 < \tan \beta < +\infty$
direction.
%
%
\subsection{THDM case}
\label{subsec:thdm_opt}

\begin{figure}
\includegraphics[width=\column]{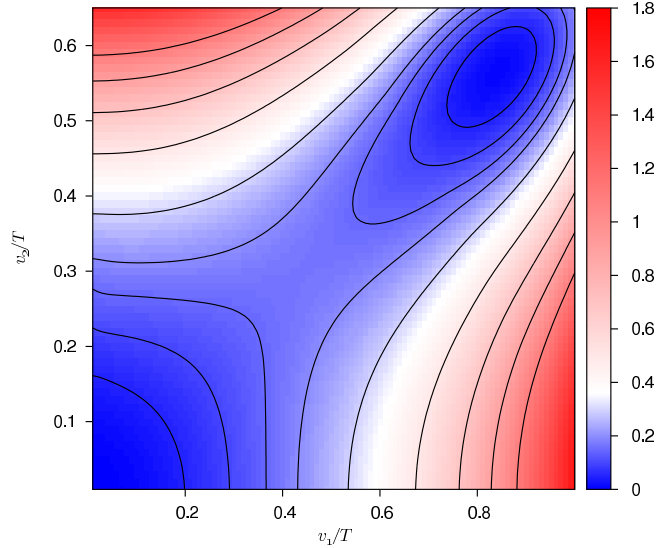}
\vskip-3mm\caption{The phase transition in the THDM type II with the
optimal parameter values. The critical temperature is $T_c =
125.39~\mathrm{GeV}$} \label{fig:1st_order_opt}
\end{figure}

Let us apply the procedure described above to the THDM type II.
In (\ref{eq:scal_pot}), there are nine parameters --
$\lambda_{1,2,3,4,5}$, $\mu_{1,2}$, $v_{01,2}$. Six of them are
free (we choose $\lambda_{1,2,3,4,5}$ and $v_{01}$). We set the
values of the couplings (listed in Appendix C), then
determine $\mu_{1,2}(T),$ and solve (\ref{eq:opt_condition}) to
obtain the $v_{01}$ value for each set of the parameters. We will
refer to the parameters obtained this way as the \textit{optimal}
parameters. The Gibbs potential with the optimal parameters taken
is shown in Fig.~\ref{fig:1st_order_opt}. The $\lambda_{1,2,3,4,5}$
values are the same as those in the case shown in Fig. \ref{fig:1st_order_no_opt},
and the jump of the order parameter is somewhat larger:
$\delta v /T_c = 1.02$.

If the parameters are optimal, and the fields $A_0$, $H^{\pm}$ are heavy
enough ($m_{H^\pm} > 80$ GeV, \cite{RPP}), the strong first-order EPT
is realized in the THDM. For small deviations from the optimal parameters,
the third Sakharov condition is still fulfilled. As a result, we see that
the baryogenesis condition is satisfied in a wide domain of the parameter
space. Five of the six free parameters of the model can be set to the
values which are constrained by modern experimental bounds and stability
requirements for the tree-level potential. Then the remaining parameter
can be calculated by using (\ref{eq:opt_condition}).
%
%
\section{Conclusions}
\label{sec:conclusions}
In the present paper we have investigated the electroweak phase transition
in the Two-Higgs Doublet Model on the base of the ring-improved one-loop
Gibbs potential. We found that there is a wide domain in the parameter
space of the model, for which the third Sakharov's baryogenesis condition
is fulfilled. The values of the parameters entering the potential
correspond to the scalar field masses that are compatible with modern
experimental constraints. The parameter values for this domain can be
found by using the introduced relation (\ref{eq:opt_condition}).

The EPT kind depends strongly on the tree-level parameter values.
We have observed that the single phase transitions, as well as the sequences
of transitions of the first and second orders, may happen.

We have concluded from our analysis that the restrictions
$\mu_1^2 = \mu_2^2$, $\lambda_1 = \lambda_2$ do not ensure that the
symmetry breaking is realized along the $\tan \beta = 1$ direction,
as it was proposed in \cite{TurokZadrozny,ClineLemieux}. We have seen
also that in the model studied in \cite{TurokZadrozny,ClineLemieux},
three degenerate vacua may coexist. Some of them can be realized as
overcooled states.

Nowadays, there are few essential experimental constraints on the THDM
parameter values: the lower bound on the mass of a charged scalar field in the
MSSM and the bounds on $\tan \beta$ (see review in \cite{RPP}). In this
situation the parameter values, for which the successful baryogenesis
is possible, could  be used as certain reference points in the study of
a model extending the MSM.

%
%
\subsubsection*{APPENDIX A\\
Information on the scalar sector}

{\footnotesize
In this appendix we present the constraints on
parameters of the scalar sector, the expressions for scalar field
masses and the relations between parameters in different
parametrizations.
%
%
\subsubsection*{\bfseries\itshape 1. Tree-level potential properties}
\label{subsec:scalar_potential_properties} Six of the nine scalar
sector parameters are free due to the minimum conditions for the
tree potential
\[
-\mu_1^2 + 4\lambda_1 v_{01}^2 + 2(\lambda_3 + \lambda_5)v_{02}^2 =
0,
\]
\begin{equation*}
-\mu_2^2 + 4\lambda_2 v_{02}^2 + 2(\lambda_3 + \lambda_5)v_{01}^2 =
0,\tag{A1}
\end{equation*}
and the VEV $v_0$ is known:
\[
v_{01}^2 + v_{02}^2 = v_0^2 = (246 \, \mathrm{GeV})^2. \nonumber
\]

For the stability of the tree-level vacuum, the potential value  at large
$v_i$ values has to be positive. This translates into Sylvester's criterion
for the quadratic form
\[
\lambda_1 v_1^4 + \lambda_2 v_2^4 + (\lambda_3 + \lambda_5)v_1^2 v_2^2. \nonumber
\]
Then the scalar field couplings are restricted to the conditions
\begin{equation*}
\label{eq:positivity_constraints} \lambda_1 > 0, \quad \lambda_2 >
0, \quad 4 \lambda_1 \lambda_2 > (\lambda_3 + \lambda_5)^2. \tag{A2}
\end{equation*}

In the present paper we  consider the case where the potential minimum
is realized at $v_{01} \neq 0$, $v_{02}\neq 0$. This is because other
cases $v_{01} \neq 0$, $v_{02}= 0$ or vice versa can be reduced to the
MSM case with several additional fields, and no mass mixing is present.
As it was shown in \cite{Land1992,Hammerschmitt_ea_1994}, the latter case 
also leads to a possibility of successful baryogenesis through a two-stage 
phase transition.

%
%
\subsubsection*{\bfseries\itshape 2. Scalar field masses}
\label{subsec:scalar_masses_properties} From (\ref{eq:scalar_basis}),
we derive the expressions for the angles $\alpha$, $\gamma$, and
$\delta$.

The $\alpha$ angle and the $h$, $H$ masses are
\[
\tan 2\alpha = \frac{A_3}{A_2 - A_1},
\]
\[
m_{h,\,H}^2 = A_1 + A_2 \mp \sqrt{(A_2 - A_1)^2 + A_3^2},
\]
\[
A_1 = -\frac{1}{2}\mu_1^2 + 6 \lambda_1 v_1^2 + (\lambda_3 +
\lambda_5) v_2^2,
\]
\[
A_2 = -\frac{1}{2}\mu_2^2 + 6 \lambda_2 v_2^2 + (\lambda_3 +
\lambda_5) v_1^2,
\]
\begin{equation*}
A_3 = 4 v_1 v_2 (\lambda_3 + \lambda_5). \tag{A3}
\end{equation*}

The $\delta$ angle and the $\chi_3$, $A_0$ masses read
\[
\tan 2\delta = \frac{B_3}{B_2 - B_1},
\]
\[
m_{\chi_3 ,\,A_0}^2 = B_1 + B_2 \mp \sqrt{(B_2 - B_1)^2 + B_3^2},
\]
\[
B_1 = -\frac{1}{2}\mu_1^2 + (2\lambda_1  + \xi (g^2 + g'^2)/8) v_1^2
+ (\lambda_4 + \lambda_5) v_2^2,
\]
\[
B_2 = -\frac{1}{2}\mu_2^2 + (2\lambda_2  + \xi (g^2 + g'^2)/8) v_2^2
+ (\lambda_4 + \lambda_5) v_1^2,
\]
\begin{equation*}
B_3 = 2 v_1 v_2 (\lambda_3 - \lambda_4 + \xi (g^2 + g'^2)/8).
\tag{A4}
\end{equation*}

The $\gamma$ angle and the $\chi^{\pm}$, $H^\pm$ masses are calculated
to be
\[
\tan 2\gamma = \frac{2 C_3}{C_2 - C_1},
\]
\[
m_{\chi_\pm ,\,H^\pm}^2 = C_1 + C_2 \mp \sqrt{(C_2 - C_1)^2 + 4
C_3^2},
\]
\[
C_1 = -\frac{1}{2}\mu_1^2 +  (2 \lambda_1 + \xi g^2/8) v_1^2 +
(\lambda_4 + \lambda_5) v_2^2,
\]
\[
C_2 = -\frac{1}{2}\mu_2^2 + (2 \lambda_2 + \xi g^2/8) v_2^2 +
(\lambda_4 + \lambda_5) v_1^2,
\]
\begin{equation*}
C_3 = v_1 v_2 (\lambda_3 + \xi g^2/8). \tag{A5}
\end{equation*}

At $v_i=v_{0i},$ the $A_0$ and $H^{\pm}$ masses are gauge-invariant,
and the Goldstone fields masses are $\sqrt{\xi} m_Z$, $\sqrt{\xi} m_W$.
The masses of the $h$, $H$, $A_0,$ and $H^{\pm}$ fields are
\[
m_h^2 = 4 \lambda_1 v_{01}^2 + 4 \lambda_2 v_{02}^2 -
\]
\[
- 4\sqrt{( \lambda_2 v_{02}^2 -  \lambda_1 v_{01}^2)^2 + (v_{01}
v_{02} (\lambda_3 + \lambda_5))^2},
\]
\[
m_H^2 = 4 \lambda_1 v_{01}^2 + 4 \lambda_2 v_{02}^2 +
\]
\[
+ 4\sqrt{( \lambda_2 v_{02}^2 -  \lambda_1 v_{01}^2)^2 + (v_{01}
v_{02} (\lambda_3 + \lambda_5))^2},
\]
\[
m_{A_0}^2 = 2(\lambda_4 - \lambda_3) (v_{01}^2 + v_{02}^2),
\]
\begin{equation*}
m_{H^{\pm}}^2 = -2\lambda_3 (v_{01}^2 + v_{02}^2). \tag{A6}
\end{equation*}
These expressions yield the following restriction on the scalar field
couplings:
\begin{equation*}
\lambda_3 < 0, \quad \lambda_4 - \lambda_3 > 0. \tag{A7}
\end{equation*}
These constraints ensure that the scalar fields $A_0$ and $H^{\pm}$
are physical ones.
%
%
\subsubsection*{\bfseries\itshape 3. Different parametrizations}
\label{subsec:different_parametrizations}

The expression for the THDM potential used in \cite{GinIvKan} is
\[
V = -\frac{1}{2}\left[ m_{11}^2 x_1 + m_{22}^2 x_2 + m_{12}^2(x_3 +
x_3^\dagger)\right] +
\]
\begin{equation*}
+ \frac{\tilde{\lambda}_1 x_1^2 + \tilde{\lambda}_2 x_2^2}{2} +
\tilde{\lambda}_3 x_1 x_2 + \tilde{\lambda}_4 x_3 x_3^\dagger +
\frac{\tilde{\lambda}_5(x_3^2 + x_3^{\dagger 2})}{2}, \tag{A8}
\end{equation*}
where
\[
 x_1 = \varphi_1^\dagger \varphi_1, \quad x_2 =
\varphi_2^\dagger \varphi_2, \quad x_3 = \varphi_1^\dagger \varphi_2
, \quad \varphi_i = \frac{1}{\sqrt{2}} \left(
\begin{array}{c}
0 \\ v_i
\end{array} \right),
\]
and we consider the CP-conserving case only. By comparing this
expression with (\ref{eq:scal_pot}) and
(\ref{eq:treelevelpotential}), we obtain the relations between the
parameter values in different parametrizations:
\[
m_{11}^2 = 2 \mu_1^2, \quad m_{22}^2 = 2 \mu_2^2, \quad
\tilde{\lambda}_1 = 8 \lambda_1, \quad \tilde{\lambda}_2 = 8
\lambda_2,
\]
\begin{equation*}
\tilde{\lambda}_3 = 4 \lambda_5, \quad \tilde{\lambda}_4 = 2
(\lambda_3 - \lambda_4), \quad \tilde{\lambda}_5 = 2 (\lambda_3 +
\lambda_4). \tag{A9}
\end{equation*}
Note that the parameter $m_{12}$ is absent in the potential investigated
in the present paper.
%
%
\subsubsection*{APPENDIX B\\
Debye masses} \label{sec:debye_masses}
Finite-temperature corrections to the $h$ and $H$ masses are given
by (\ref{eq:debye_masses_h_H}). The Debye masses of the
$\chi^{\pm}$, $H^{\pm}$, $\chi_3$, and $A_0$ fields are calculated to be
\[
\delta m^2_{\chi^{\pm}} = T^2 \left((2\lambda_1 + \frac{1}{2} G_t^2)
\cos^2 \gamma + (2\lambda_2 + \frac{1}{2} G_b^2) \sin^2 \gamma +
\right.
\]
\[
\left. + \frac{\lambda_3 + \lambda_4 + 4\lambda_5}{6} + \frac{3g^2 +
g'^2}{16} \right),
\]
\[
\delta m^2_{H^{\pm}} = T^2 \left((2\lambda_1 + \frac{1}{2} G_t^2)
\sin^2 \gamma + (2\lambda_2 + \frac{1}{2} G_b^2) \cos^2 \gamma +
\right.
\]
\[
\left. + \frac{\lambda_3 + \lambda_4 + 4\lambda_5}{6} + \frac{3g^2 +
g'^2}{16} \right),
\]
\[
\delta m^2_{\chi_3} = T^2 \left((2\lambda_1 + \frac{1}{2} G_t^2)
\cos^2 \delta + (2\lambda_2 + \frac{1}{2} G_b^2) \sin^2 \delta +
\right.
\]
\[
\left. + \frac{\lambda_3 + \lambda_4 + 4\lambda_5}{6} + \frac{3g^2 +
g'^2}{16} \right),
\]
\[
\delta m^2_{A_0} = T^2 \left((2\lambda_1 + \frac{1}{2} G_t^2) \sin^2
\delta + (2\lambda_2 + \frac{1}{2} G_b^2) \cos^2 \delta + \right.
\]
\begin{equation*}
\label{eq:debye_masses_app} \left. + \frac{\lambda_3 + \lambda_4 +
4\lambda_5}{6} + \frac{3g^2 + g'^2}{16} \right). \tag{B1}
\end{equation*}
For the THDM type I and for the doublet-universal parametrization, we
have the following differences:

1. THDM type I. The $G_b$ coupling is always small as compared to
the $G_t$ coupling and can be neglected.

2. The doublet-universal parametrization. In this case, $G_b$ can
be neglected. Also, one has to make substitutions into
(\ref{eq:debye_masses_app}) using the prescription
\begin{eqnarray}
\lefteqn{
G_t^2 \sin^2 \alpha,\gamma,\delta \to G_t^2(\sin \alpha,\gamma,\delta + \cos \alpha,\gamma,\delta)^2,
}\nonumber\\
\lefteqn{
G_t^2 \cos^2 \alpha,\gamma,\delta \to G_t^2(\sin \alpha,\gamma,\delta - \cos \alpha,\gamma,\delta)^2.
}\nonumber
\end{eqnarray}

We present the contributions to the gauge boson Debye mass from
different sectors of the model. The $Z$ boson Debye mass is
\begin{eqnarray}
\lefteqn{
\delta m^2_Z = \delta m^{s\,2}_Z + \delta m^{f\,2}_Z + \delta m^{g\,2}_Z,
}\nonumber\\
\lefteqn{
\delta m^{s\,2}_Z = \frac{T^2}{3} \frac{g^4 + g'^4}{g^2 + g'^2}, \quad
\delta m^{f\,2}_Z = \frac{T^2}{12} \frac{3 g^4 + 29 g'^4}{g^2 + g'^2},
}\nonumber\\
\lefteqn{
\delta m^{g\,2}_Z = \frac{2}{3} T^2 \frac{g^4}{g^2 + g'^2}. \nonumber
}
\end{eqnarray}
For the $W$ boson mass, we have
\begin{eqnarray}
\lefteqn{
\delta m^2_W = \delta m^{s\,2}_W + \delta m^{f\,2}_W + \delta m^{g\,2}_W,
}\nonumber\\
\lefteqn{
\delta m^{s\,2}_W = \frac{1}{3}g^2 T^2, \quad \delta m^{f\,2}_W = g^2 T^2, \quad \delta m^{g\,2}_W = \frac{2}{3}g^2 T^2.
}\nonumber
\end{eqnarray}
The $s$, $f$, and $g$ superscripts denote the contributions coming from
the scalar, fermion, and gauge boson sectors, respectively.
%

%
\subsubsection*{APPENDIX C\\
Parameter values for figures}
\label{sec:parameter_values_for_figures} In this appendix, we adduce
the parameter values  used in the calculations. The $t$ and
$b$ mass values were taken 175 GeV and 4.2 GeV, respectively. }

\begin{table}[h]
\noindent \caption{The parameter values for (\ref{eq:scal_pot}) used
for plotting the Gibbs potential}\vskip3mm\tabcolsep6.9pt \noindent
{\footnotesize
\begin{tabular}{ccccccc}
\hline
\multicolumn{1}{c}{fig.} & \multicolumn{1}{|c}{$\lambda_1$} & \multicolumn{1}{|c}{$\lambda_2$} & \multicolumn{1}{|c}{$\lambda_3$} & \multicolumn{1}{|c}{$\lambda_4$} & \multicolumn{1}{|c}{$\lambda_5$} & \multicolumn{1}{|c}{$v_{01}/v_{02}$}\\
\hline
\ref{fig:1st_order_no_opt} & 0.045 & 0.135 & --0.27 & 0.675 & 0.27 & 1.91\\
\ref{fig:1st_order_sequence} & 0.045 & 0.135 & --0.27 & 0.315 & 0.27 & 2\\
\ref{fig:1st2nd_order_sequence_1} & 0.095 & 0.2375 & --0.855 & 0.655 & 0.855 & 2.64\\
\ref{fig:1st_order_sequence_Cline} & 0.16 & 0.16 & --0.52 & 0 & 0.72 & 1\\
\ref{fig:1st_order_opt} & 0.045 & 0.135 & --0.27 & 0.675 & 0.27 & 1.86\\
\hline
\end{tabular}}
\end{table}

\begin{table}[h]
\noindent \caption{The parameter values resulting in a sequence of
second-order phase transitions}\vskip3mm\tabcolsep10.2pt
\noindent {\footnotesize
\begin{tabular}{cccccc}
\hline
\multicolumn{1}{c}{$\lambda_1$} & \multicolumn{1}{|c}{$\lambda_2$} & \multicolumn{1}{|c}{$\lambda_3$} & \multicolumn{1}{|c}{$\lambda_4$} & \multicolumn{1}{|c}{$\lambda_5$} & \multicolumn{1}{|c}{$v_{01}/v_{02}$}\\
\hline
0.045 & 0.135 & --0.27 & 0.675 & 0.27 & 1.22\\
\hline
\end{tabular}}
\end{table}

\end{document}